\documentclass[twocolumn,preprintnumbers,amsmath,amssymb,aps,prb,nopacs]{revtex4}
\usepackage{graphicx}
\begin{document}

\title{
Magnus-Induced Ratchet Effects For Skyrmions Interacting with Asymmetric Substrates} 
\author{
C. Reichhardt, D. Ray,  and  C. J. Olson Reichhardt$^*$ 
}
\affiliation{
Theoretical Division and Center for Nonlinear Studies,
Los Alamos National Laboratory, Los Alamos, New Mexico 87545 USA
}

\date{\today}

\begin{abstract}
 When a particle is driven with an ac force over an asymmetric
 potential, it can undergo a ratchet effect that produces a net dc
 motion of the particle.
  Ratchet effects have been observed in numerous systems such as
  superconducting vortices on asymmetric pinning substrates.
  Skyrmions, stable topological spin textures with particle-like properties,
  have many similarities to vortices but their behavior is strongly
  influenced by
  non-disspative effects arising from a Magnus term in their equation of
  motion.  
  We show using numerical simulations
  that pronounced ratchet effects can occur for ac driven skyrmions
  moving over asymmetric quasi-one-dimensional substrates.  We find a
  new type of ratchet effect called a Magnus-induced transverse ratchet that
  arises when the ac driving force is applied perpendicular rather than
  parallel to the asymmetry direction of the substrate.
  This transverse ratchet effect only occurs when the Magnus term is finite,
  and the threshold ac amplitude needed to induce it decreases as the Magnus
  term becomes more prominent.  Ratcheting skyrmions follow
  ordered orbits in which the net displacement parallel to the
  substrate asymmetry
  direction is quantized.
  Skyrmion ratchets represent a new ac current-based method for controlling
  skyrmion positions and motion for spintronic applications.
  
\end{abstract}
\maketitle

Ratchet effects can arise when particles placed in an
asymmetric environment are subjected to
suitable nonequilbrium conditions
such as an externally applied ac drive,
which produces a net dc motion of the particles \cite{1}.
Ratchet effects have been realized for colloidal systems \cite{2},
cold atoms in optical traps \cite{3},
granular media on sawtooth substrates
\cite{4}, and the motion of cells crawling on patterned asymmetric
substrates\cite{5}, and they 
can be exploited to create devices such as
shift registers for domain walls moving in an asymmetric
substrate \cite{6}.
There has been intense study of 
ratchet effects 
for magnetic flux vortices in type-II superconductors
interacting with quasi-one-dimensional (q1D) or two-dimensional (2D)
asymmetric pinning substrates.
The vortices can be effectively described as overdamped particles
moving over an asymmetric substrate, and
under ratcheting conditions
the application of an
ac driving current produces a net dc flux flow \cite{7,8,9,10,11,12}.
The simplest
vortex ratchet geometry was first proposed by Lee {\it et al.} \cite{7},
where an effectively 2D assembly of vortices was driven over
a q1D asymmetrically modulated substrate using an ac driving force oriented
along the direction of the substrate asymmetry.

Skyrmions in chiral magnets, initially discovered in 
MnSi \cite{13} and subsequently identified in numerous other materials
at low \cite{14,15,16,17} and room temperatures \cite{18,19,20},
have many similarities to superconducting
vortices.
Skyrmions are spin textures forming topologically stable 
particle-like objects \cite{17} that can be set into motion
by the application of a spin-polarized current \cite{21,22,23,24,25}.
As a function of the external driving,
skyrmions can exhibit a depinning transition similar to
that found for vortices, and from transport measurements it is possible to 
construct skyrmion velocity versus applied force curves
\cite{22,24,25,26,27,liang}.  
It has been shown that the dynamics of skyrmions interacting with pinning
can be captured by an effective particle equation of  
motion, which produces pinning-depinning and transport properties
that agree with those obtained using a continuum-based model \cite{26}, 
just as a particle-based description 
for vortices can effectively capture the vortex dynamics
in the presence of pinning sites \cite{28}. 
Due to their size scale and the low currents required to move them,
skyrmions show tremendous promise for applications in
spintronics
such as race-track type memory devices
\cite{29,30} originally proposed for
domain walls \cite{31},
and it may be possible to create skyrmion logic
devices similar to those
that harness magnetic domain walls \cite{32}.
In order to realize such skyrmion applications, new methods
must be developed to precisely control skyrmion positions and
motion.

The most straightforward approach to a skyrmion ratchet is to utilize
the same types of asymmetric substrates known to produce ratchet effects
in vortex systems.
There are, however, 
important differences between skyrmion and vortex dynamics due to
the strong 
non-dissipative Magnus component of skyrmion motion
\cite{13,17,22,24,26,27} that is absent in the vortex system.
The Magnus term produces a skyrmion velocity contribution
that is perpendicular to the direction of an applied external force.
In the presence of pinning, the Magnus
term reduces the effectiveness of the pinning by causing the skyrmions to
deflect around the edges of
attractive pinning sites \cite{24,26,27} rather than being
captured by the pinning sites as overdamped vortices are.
The Magnus term also produces complex winding orbits for skyrmions moving
in confined regions \cite{33} or through pinning sites
\cite{24,26,27,34,35,36}.
The ratio
of the strength of the Magnus term $\alpha_m$ to the dissipative
term $\alpha_{d}$ can be 10 or higher for skyrmion systems \cite{17}.
In contrast, although it is possible for a Magnus effect to appear for
vortices in superconductors, it is generally very weak
so that the vortex dynamics is dominated by dissipation \cite{28}.    
A key question is how the Magnus force could impact
possible ratchet effects for skyrmions, and
whether new types of ratchet phenomena can be realized for skyrmion
systems that are not accessible in overdamped systems \cite{37}.

\begin{figure}
  \includegraphics[width=3.5in]{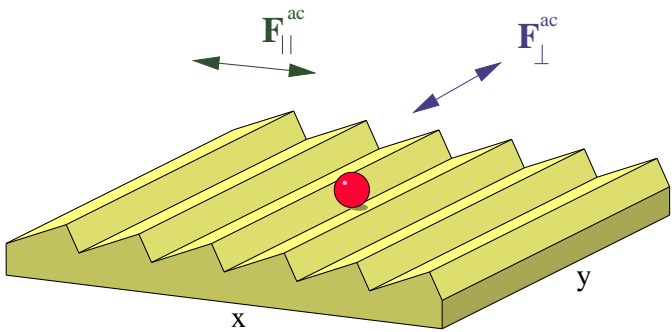}
\caption{
  {\bf Asymmetric substrate geometry}:
  A schematic of the sample geometry where a skyrmion is
  placed on a q1D asymmetric periodic substrate.  An ac driving force
  $F^{ac}\cos(\omega t)$ is applied parallel ($F^{ac}_{\parallel}$) or
  perpendicular ($F^{ac}_{\perp}$) to the substrate asymmetry direction.
}
\label{fig:1}
\end{figure}

In this work we numerically examine a 2D assembly of 
skyrmions interacting with a q1D asymmetric periodic substrate shown
schematically in Fig.~1.
An ac drive is applied
either parallel ($F^{ac}_{\parallel}$) or perpendicular
($F^{ac}_{\perp}$)
to the substrate asymmetry direction, which
runs along the $x$ axis.
In the overdamped limit, no ratchet effect appears for
$F^{ac}_{\perp}$; however,
we find that the Magnus effect produces a novel ratchet
effect for $F^{ac}_{\perp}$ when it curves the skyrmion
trajectories into the asymmetry or $x$ direction.
We model 
the skyrmion dynamics in a sample with periodic
boundary conditions using a particle based description
\cite{26,36} and vary the ratio of the
Magnus to dissipative dynamic terms, the amplitude and frequency of the ac
drive,
and the strength of the substrate.

\section*{Simulation}
The equation of motion for a single skyrmion
with velocity ${\bf v}$ moving in the $x-y$ plane
is  
\begin{equation}
  \alpha_{d}{\bf v} + \alpha_{m}{\hat z}\times {\bf v} = {\bf F}^{SP}
  + {\bf F}^{ac}_{\parallel,\perp}
\end{equation}
The damping term $\alpha_{d}$
keeps the skyrmion velocity aligned with the
direction of the net external force,
while the Magnus term $\alpha_{m}$
rotates the velocity toward the direction perpendicular to the
net external forces.
To examine the role of the Magnus term
we vary the relative strength
$\alpha_{m}/\alpha_{d}$
of the Magnus term to the dissipative term
under the constraint 
$\alpha^2_{d} + \alpha^{2}_{m} = 1$,
which maintains a constant magnitude of the skyrmion velocity.
The substrate force 
${\bf F}^{SP}  = -\nabla U(x) {\bf \hat x}$ arises from a ratchet potential
\begin{equation}
U(x)  = U_{0}[\sin(2\pi x/a) + 0.25\sin(4\pi x/a)],
\end{equation}
where $a$ is the periodicity of the substrate and  
we define the strength of the substrate to be $A_{p} = 2\pi U_{0}/a$.
The ac driving 
term is
either ${\bf F}^{ac}_{\parallel} = F^{ac}_{\parallel}\cos(\omega t){\bf \hat x}$ or 
${\bf F}^{ac}_{\perp} = F^{ac}_{\perp}\cos(\omega t){\bf \hat y}$. 
We measure the time-averaged skyrmion velocities
parallel
$\langle V\rangle_\parallel\equiv
2\pi\langle {\bf v} \cdot {\bf \hat x} \rangle/\omega a$
or perpendicular
$\langle V\rangle_\perp\equiv
2\pi\langle {\bf v} \cdot {\bf \hat y} \rangle/\omega a$
to the substrate asymmetry direction
as we vary the ac amplitude, substrate strength, or $\omega$. 
Here we focus primarily on samples with $A_{p} = 1.5$
and $\omega = 2.5\times 10^{-6}$ inverse simulation time steps, and
we consider
the sparse limit of a single skyrmion
which can also apply to regimes in which skyrmion-skyrmion
interactions are negligible.         

\section*{Results and Discussion}

\begin{figure}
\includegraphics[width=3.5in]{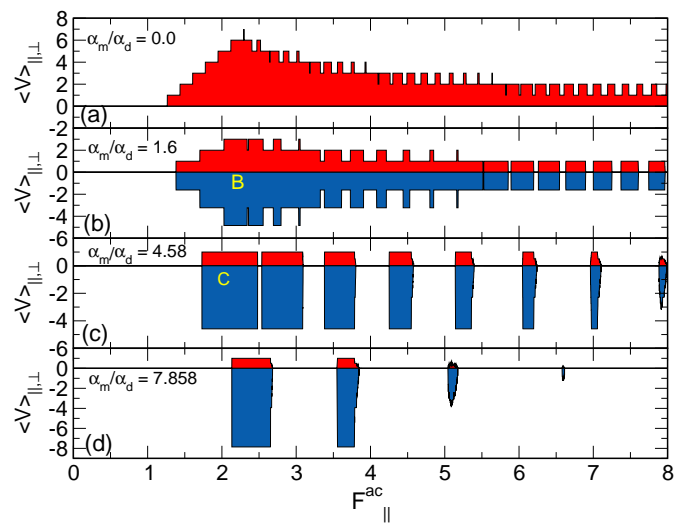}
\caption{
  {\bf Parallel and perpendicular ratchet velocities for parallel drive}:
  The skyrmion velocity parallel $\langle V\rangle_{\parallel}$ (red)
  and perpendicular $\langle V\rangle_{\perp}$ (blue) to the
  substrate asymmetry vs
  $F^{ac}_{\parallel}$ for a substrate strength
  of $A_{p} = 1.5$.
  (a) At $\alpha_{m}/\alpha_{d} = 0.0$, the overdamped limit,
  only $\langle V\rangle_{\parallel}$ is nonzero and shows quantized steps
  with 
  $\langle V\rangle_{\parallel} = n$, where $n$ is an integer.
  (b) At $\alpha_{m}/\alpha_{d} = 1.6$, there is
  quantized motion in both the parallel and perpendicular directions.
  (c) $\alpha_{m}/\alpha_{d} = 4.58$. (d) $\alpha_{m}/\alpha_{d} = 7.85$. 
  The points labeled B and C indicate where 
  the skyrmion orbits in Fig.~3(b,c) were obtained.
}
\label{fig:2}
\end{figure}

\subsection{ac Driving Parallel to Substrate Asymmetry}
We first apply the ac driving force parallel to
the asymmetry direction, a configuration previously studied for
the same type of substrate in vortex systems \cite{7,12}. 
In Fig.~2 we plot $\langle V\rangle_{\parallel}$ and
$\langle V\rangle_{\perp}$ versus $F^{ac}_{\parallel}$ for samples with
$A_{p} = 1.5$, so that the maximum magnitude of the substrate force
is $F^{s}_{\rm max}=2.0$ in the negative $x$ direction and
$F^s_{\rm max}=1.0$ in the positive $x$ direction.
The overdamped limit $\alpha_{m}/\alpha_{d} = 0.0$ appears in
Fig.~2(a),
where $\langle V\rangle_{\perp} = 0.0$
and there is a ratchet effect only along the parallel or $x$-direction 
for $F^{ac}_{\parallel} > 1.25$.
The skyrmion velocity is quantized and forms a series of steps 
with $\langle V\rangle_{\parallel}= n$ where $n$ is an integer.
On the highest step in Fig.~2(a) near $F^{ac}_{\parallel}=2.3$, $n = 7$.
The quantization indicates that during one ac driving period,
the skyrmion moves a net distance of $na$ in the parallel direction. 
For very large values of $F^{ac}_{\parallel}$, the velocity quantization is
lost and the parallel ratchet effect gradually diminishes.
In Fig.~2(b) we plot $\langle V\rangle_{\parallel}$ and
$\langle V\rangle_{\perp}$ for a sample with
$\alpha_{m}/\alpha_{d} = 1.6$,
showing a nonzero skyrmion velocity component for both the parallel and
perpendicular directions.
There is again a quantization of the parallel velocity
$\langle V\rangle_{\parallel}$, and the Magnus term
transfers this quantization
into the perpendicular direction even though there is no periodicity of
the substrate along the $y$ direction.
Thus, we find $\langle V\rangle_{\perp}=n\alpha_m/\alpha_d$.
For $F^{ac}_{\parallel} > 5.0$, we find windows in which
the ratcheting effect is lost and
$\langle V\rangle_{\parallel}=\langle V\rangle_{\perp}=0$.
As $\alpha_{m}/\alpha_{d}$ increases, the maximum value
of $n$ decreases,
so that
at $\alpha_{m}/\alpha_{d} = 4.58$ in Fig.~2(c),
only the $n = 1$ step appears.
At the same time, the width of the non-ratcheting
windows increases,
as shown
in Fig.~2(d) for $\alpha_{m}/\alpha_{d} = 7.858$.
These
results show that while a ratchet effect does
occur for skyrmion systems, an increase in the Magnus term
decreases the range of ac drives over which the ratchet
effect can be observed.

\begin{figure}
  \includegraphics[width=3.5in]{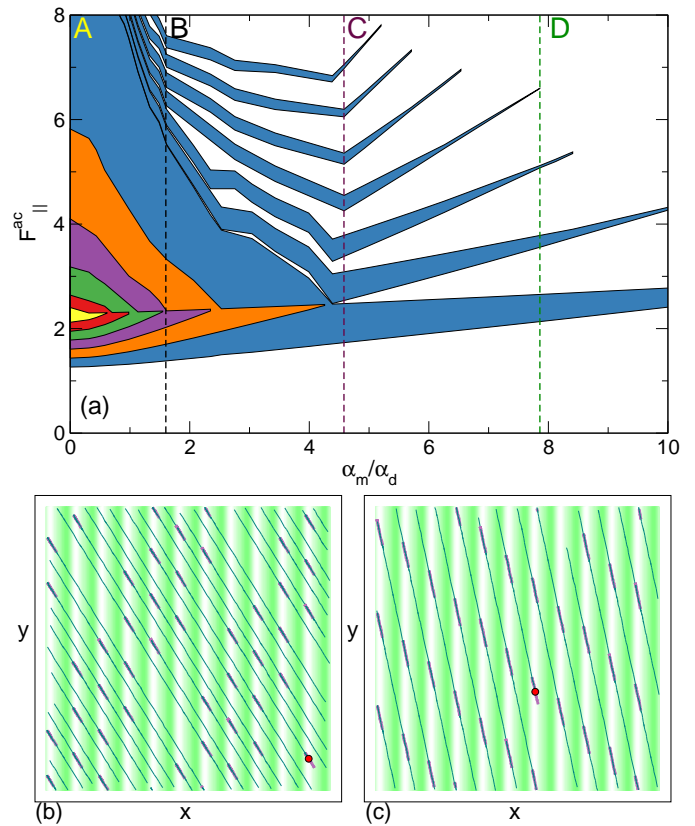}
  \caption{
    {\bf Ratchet region envelopes and example ratchet orbits for
      parallel ac driving}:
  ({\bf a}) A plot of $F^{ac}_{\parallel}$ vs $\alpha_m/\alpha_d$
  indicating all $n=1$ (blue) ratcheting regions, and
  the inner envelopes of the
  $n=2$ (orange), $3$ (purple), $4$ (green), $5$ (red) 
  and $6$ (yellow) ratcheting regions.
  The boundaries of the inner envelopes are defined as the
  point at which the ratchet velocity first reaches the
  value $n$ followed by the point at which it first drops
  below $n$.
  The y-axis marked $A$ and the dashed lines marked $B-D$ are the
  values of $\alpha_m/\alpha_d$ for which
  the velocity curves in Fig.~2 were obtained.
  The minimum ac amplitude required to produce a ratchet 
  effect increases with increasing $\alpha_{m}/\alpha_{d}$.
  ({\bf b,c}) Skyrmion trajectory images.
  White (green) regions are high (low) areas
  of the substrate potential.  Thin black lines indicate forward motion
  of the skyrmion and thick purple lines indicate backward motion.
  The red circle is the skyrmion.
  ({\bf b})
  The $n = 3$ orbit at $\alpha_{m}/\alpha_{d} =  1.6$
  for $F^{ac}_{\parallel} = 2.2$ from Fig.~2(b).
  The skyrmion moves in straight lines
  and translates $3a$ in the $x$ direction during one ac drive cycle.
  ({\bf c}) The $n = 1$ orbit at $\alpha_{m}/\alpha_{d} = 4.58$
  for $F^{ac}_{\parallel} = 2.0$ from 
  Fig.~2(c), where the skyrmion translates a distance $a$
  in the $x$ direction during every ac drive cycle.     
}
\label{fig:3}
\end{figure}

By lowering the ac frequency $\omega$ or increasing the substrate
strength $A_p$,
it is possible to increase the range and magnitude
of the skyrmion ratchet effect.
In  Fig.~3(b,c) we show representative skyrmion 
orbits for the system in Fig.~2(b,c).
Figure~3(b) shows the orbits at $\alpha_{m}/\alpha_{d} = 1.6$ for
$F^{ac}_{\parallel} = 2.2$, corresponding to the $n = 3$ step
in Fig.~2(b). Here the skyrmion moves along straight lines oriented at an
angle to the direction of the applied ac drive.
During each ac cycle, the skyrmion first translates
a distance $3a$ in the positive $x$ direction and then travels in the
reverse direction for a distance of $a/2$ before striking the top of the
potential barrier, which it cannot overcome in the reverse $x$ direction for
this magnitude of ac drive.  It repeats this motion in each cycle.
In Fig.~3(c) we show the skyrmion orbit at $\alpha_{m}/\alpha_{d} = 4.58$
for $F^{ac}_{\parallel} = 2.0$, corresponding 
to the $n = 1$ orbit  in Fig.~2(c).  The angle the skyrmion motion makes with
the ac driving direction is much steeper, and the skyrmion is displaced a net
distance of $a$ in the $x$ direction during each ac driving cycle.

In order to clarify the evolution of the ratchet phases as a function 
of $\alpha_{m}/\alpha_{d}$ and $F^{ac}_{\parallel}$,
in Fig.~3(a) we plot the ratchet envelopes for the $n=1$ to $n=6$ steps.
For the $n=1$ case we show, in blue, all of the regions in the
$F^{ac}_{\parallel} - \alpha_m/\alpha_d$
plane where $n=1$ steps appear.
For $n=2$ to $n=6$, to prevent overcrowding of the graph,
instead of plotting all ratchet regions we show only
the inner step envelope for each $n=n_i$, obtained from simulations in which
$F^{ac}_{\parallel}$ is swept up from zero, and defined 
to extend from the drive at which $n$ first reaches $n_i$ to the
drive at which $n$ first drops below $n_i$.
We find that the minimum $F^{ac}_{\parallel}$ required to induce a ratchet
effect increases linearly with increasing
$\alpha_m/\alpha_d$, and that the
ratcheting regions form a series of tongue features
(shown for $n=1$ only). 
The ratchet effects persist up to and beyond $\alpha_m/\alpha_{d} = 10$.
For $\alpha_{m}/\alpha_{d} = 0$,
the strongest ratchet effects with the largest values of $n$
occur near $F^{ac}_{\parallel} = 2.3$,
corresponding to a driving force that is slightly higher than the
maximum force exerted on the skyrmion by the substrate when it is moving
with a negative $x$ velocity component.
As $F^{ac}_{\parallel}$ increases above this value, the skyrmion
can slip backward over more than one substrate plaquette during each
ac cycle, limiting its net forward progress.

\begin{figure}
\includegraphics[width=3.5in]{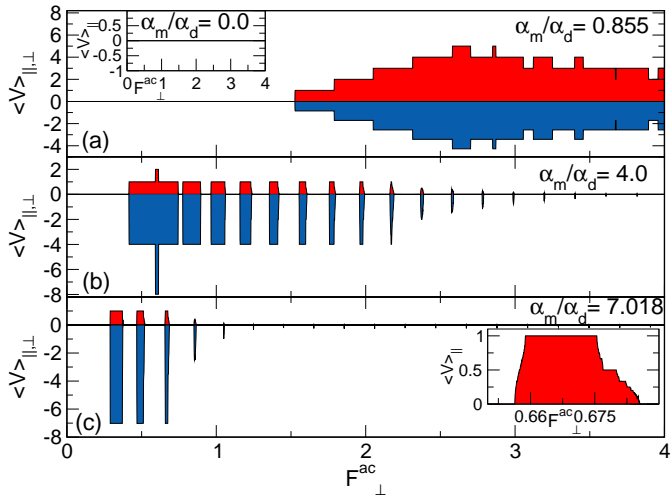}
\caption{
  {\bf Parallel and perpendicular ratchet velocities for perpendicular
    drive}:
  The skyrmion velocity parallel $\langle V\rangle_{\parallel}$ (red)
  and perpendicular $\langle V\rangle_{\perp}$ (blue) to the substrate
  asymmetry vs $F^{ac}_{\perp}$ for samples with $A_{p} = 1.5$.
  ({\bf a})
  At $\alpha_{m}/\alpha_{d} = 0.855$, ratcheting occurs in both the
  parallel and perpendicular directions for $F^{ac}_{\perp} > 1.5$, and 
  $\langle V\rangle_{\parallel} = n$ with integer $n$.
  Inset: At $\alpha_{m}/\alpha_{d} = 0$, there is no ratchet effect.
  ({\bf b}) $\alpha_{m}/\alpha_{d} = 4.0$.
  ({\bf c}) $\alpha_{m}/\alpha_{d} = 7.018$. Inset:
  A blowup of $\langle V\rangle_{\parallel}$ from the main panel
  at the third $n=1$ step showing that additional fractional steps
  appear at velocity values such as $n/m = 1/2$. 
}
\label{fig:4}
\end{figure}

\subsection{ac Driving Perpendicular to Substrate Asymmetry}
In Fig.~4 we plot $\langle V\rangle_{\parallel}$ and
$\langle V\rangle_{\perp}$ versus $F^{ac}_{\perp}$
for $A_{p} = 1.5$. The inset in Fig.~4(a) shows 
that for
$\alpha_{m}/\alpha_{d} = 0$,
$\langle V\rangle_{\parallel} = \langle V\rangle_{\perp} = 0$,
indicating that there is no ratchet effect in either direction. 
When the Magnus term is finite, as in Fig.~4(a) where
$\alpha_{m}/\alpha_{d} = 0.855$, 
pronounced ratcheting occurs in both the parallel and perpendicular
directions.
We describe this as a Magnus-induced transverse ratchet effect.
Here $\langle V\rangle_{\parallel}$ is quantized at integer values,
with the largest step in Fig.~4(a) falling at $n=5$, and
$\langle V\rangle_{\perp}/\langle V\rangle_{\parallel} = \alpha_{m}/\alpha_{d}$.
As $\alpha_{m}/\alpha_{d}$ increases, Fig.~4(b,c) shows that the maximum value
of $F^{ac}_{\perp}$ at which ratcheting occurs decreases, as does
the maximum value of $n$ and the widths of the ratcheting windows.
In addition to the integer steps in $\langle V\rangle_{\parallel}$, we
also observe fractional steps, as shown in the inset of
Fig.~4(c) for $\alpha_m/\alpha_d=7.018$, where we highlight
the third $n = 1$ step.
On the upper $n=1$ step edge there is a plateau at 
$\langle V\rangle_{\parallel} = 1/2$ and additional steps at
$\langle V\rangle_{\parallel}=3/4$, 2/3, 1/3, and $1/4$, which are accompanied by
even smaller
steps that form a devil's staircase structure \cite{devil}.
For higher values of $F^{ac}_{\perp}$, smaller plateaus
appear at rational fractional velocity values
$\langle V\rangle_{\parallel}=n/m$ with integer $n$ and $m$,
while full locking to the integer $n = 1$
step no longer occurs.
The $n/m$ rational locking steps also occur for parallel ac driving
but are much weaker than for perpendicular ac driving.

\begin{figure}
\includegraphics[width=3.5in]{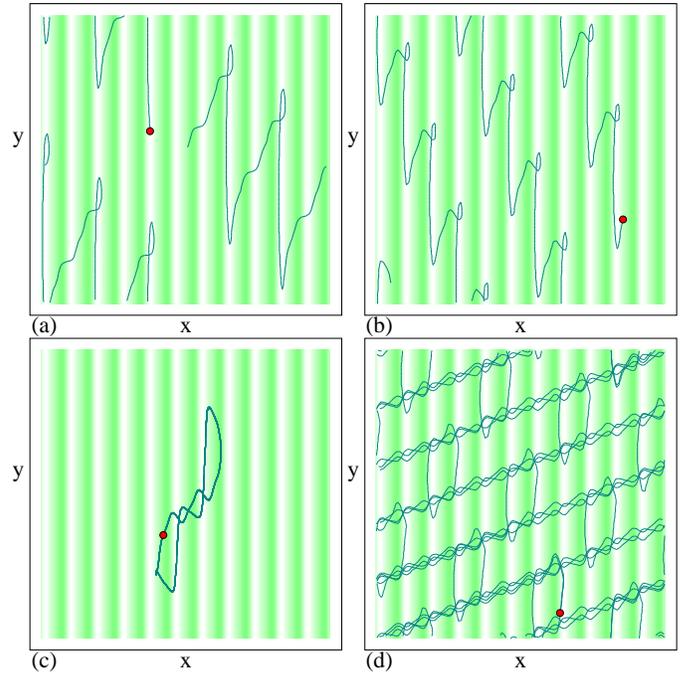}
\caption{
  {\bf Example ratchet orbits for perpendicular ac driving}:
  Skyrmion trajectory images from the system in Fig.~4.
  White (green) regions are high (low) areas of
  the substrate potential.  Black lines indicate the motion of the skyrmion;
  the red circle is the skyrmion.
  ({\bf a}) The $n = 2$ step at $\alpha_m/\alpha_{d}  = 0.855$
  for $F^{ac}_{\perp} = 2.0$.
  ({\bf b}) The $n = 1$ step at $\alpha_{m}/\alpha_{d} = 4.0$ for
  $F^{ac}_{\perp} = 0.7$.
  ({\bf c}) A non-ratcheting orbit at $\alpha_{m}/\alpha_{d} = 4.0$
  for $F^{ac}_{\perp} = 0.76$.
  ({\bf d}) For higher ac drives the orbits become more
  complicated, as shown here for $\alpha_{m}/\alpha_{d} = 4.0$ at
  $F^{ac}_{\perp} = 1.97$.  
}
\label{fig:5}
\end{figure}

The skyrmion orbits for perpendicular ac driving 
differ markedly from those that appear under parallel ac driving,
since the Magnus term induces a velocity 
component perpendicular to the direction of the external driving force.
As a result, perpendicular ac drives induce parallel skyrmion motion
that interacts with the substrate asymmetry.
Figure~5(a) shows a skyrmion orbit on the $\langle V\rangle_{\parallel}= 2$ step 
for $F^{ac}_{\perp} = 2.0$ and $\alpha_{m}/\alpha_{d} = 0.855$
from Fig.~4(a). 
The skyrmion no longer moves in straight lines, as it did for the
parallel ac drive in Fig.~3.  Instead it follows a complicated 2D
trajectory, translating a net distance of $2a$ in the $x$ direction
during every ac drive cycle.
In Fig.~5(b) we plot an $n = 1$ orbit at $F^{ac}_{\perp} = 0.7$ for
$\alpha_{m}/\alpha_{d} = 4.0$ from Fig.~4(b),  
where there is a similar orbit structure in which
the skyrmion translates a net distance of $a$ in the $x$ direction in each
ac drive cycle.  
Figure~5(c) shows a non-ratcheting orbit for 
$\alpha_{m}/\alpha_{d} = 4.0$ at $F^{ac}_{\perp} = 0.75$,
where the skyrmion traces out a complicated periodic closed cycle.
For higher ac amplitudes, the ratcheting orbits become
increasingly intricate, as shown in Fig.~5(d) for
$\alpha_{m}/\alpha_{d} = 4.0$ at  
$F^{ac}_{\perp} = 1.97$.

\begin{figure}
  \includegraphics[width=3.5in]{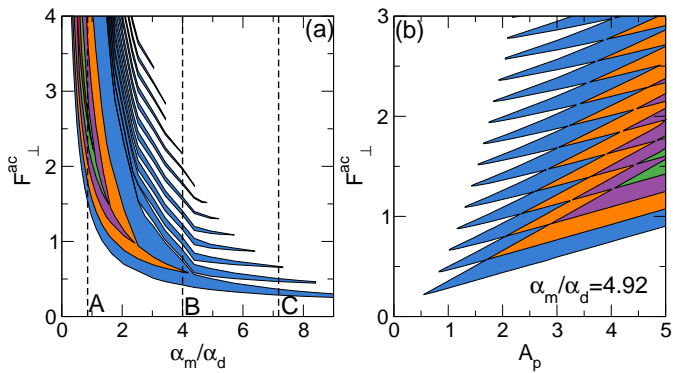}
  \caption{
    {\bf Ratchet region envelopes for perpendicular ac driving}:
    ({\bf a}) A plot of $F^{ac}_{\perp}$ vs $\alpha_m/\alpha_{d}$
    indicating all $n=1$ (blue) ratcheting regions, and the
    inner envelopes of the $n=2$ (orange), 3 (purple), 4 (green), and 5 (red)
    ratcheting regions.
    The dashed lines marked $A-C$ are the values of $\alpha_m/\alpha_d$
    for which the velocity curves in Fig.~4 were obtained.
    There is no ratchet effect at the overdamped limit of
    $\alpha_m/\alpha_{d} = 0$, and the minimum ac amplitude required to
    produce a ratchet effect decreases with increasing $\alpha_{m}/\alpha_{d}$.
    ({\bf b}) A plot of $F^{ac}_{\perp}$ vs $A_{p}$
    for $\alpha_{m}/\alpha_{d} = 4.92$ indicating the ratcheting regions
    with $n=1$ (blue), 2 (orange), 3 (purple), and 4 (green).
    As the substrate strength increases, stronger ratchet effects appear. 
}
\label{fig:6}
\end{figure}
  
In Fig.~6(a) we show the full $n=1$ ratchet regions and the inner envelopes
of the $n=2$ to 5 ratchet regions as a function of
$F^{ac}_{\perp}$ and $\alpha_{m}/\alpha_{d}$.
For low values of  $\alpha_m/\alpha_{d}$,
there is no ratcheting since the Magnus term is too weak
to bend the skyrmion trajectories
far enough to run along the direction parallel to the asymmetry.
The minimum ac amplitude required to generate a ratchet effect decreases 
as the strength of the Magnus term increases,
opposite to what we found for parallel ac driving
in Fig.~3(a).
Here, the skyrmion trajectories are more strongly curved into
the parallel direction as $\alpha_{m}/\alpha_{d}$ increases,
so that a lower ac amplitude is needed to push the skyrmions over the
substrate potential barriers.
In contrast, for a parallel ac drive
the Magnus term curves the skyrmion trajectories out of the parallel
direction, so that a larger ac amplitude must be applied for the
skyrmions to hop over the substrate maxima.
We also find that as $\alpha_m/\alpha_{d}$ increases,
the minimum ac force needed to produce a ratchet effect
drops below the pinning force exerted by the substrate.
This cannot occur in overdamped systems, and is an indication that
the Magnus term induces some inertia-like behavior.
The threshold ac force value $F^{th}$ for ratcheting to occur can be fit to
$F^{th}_{\perp} \propto (\alpha_{m}/\alpha_{d})^{-1}$ for perpendicular driving,
while for parallel driving,
$F^{th}_{\parallel} \propto \alpha_{m}/\alpha_{d}$. 
As $\alpha_{m}/\alpha_{d}$ increases, the extent of the 
regions where ratcheting occurs decreases; however,
by increasing the substrate strength, the regions of ratcheting broaden
in extent and the magnitude of the ratchet effect increases.
In Fig.~6(b) we show the evolution of the $n = 1$, 2, 3 and $4$   
ratchet regions as a function of $F^{ac}_{\perp}$ and $A_{p}$
for fixed $\alpha_{m}/\alpha_{d} = 4.92$. 
Here, as $A_p$ increases, the magnitude of the
ratchet effect
increases, forming a
series of tongues.
The minimum ac force needed to induce a ratchet effect increases linearly with
increasing $A_{p}$.
These results indicate that skyrmions can exhibit a new type of
Magnus-induced ratchet effect that appears
when the driving force is applied perpendicular to the asymmetry
direction of the substrate.  

\begin{figure}
\includegraphics[width=3.5in]{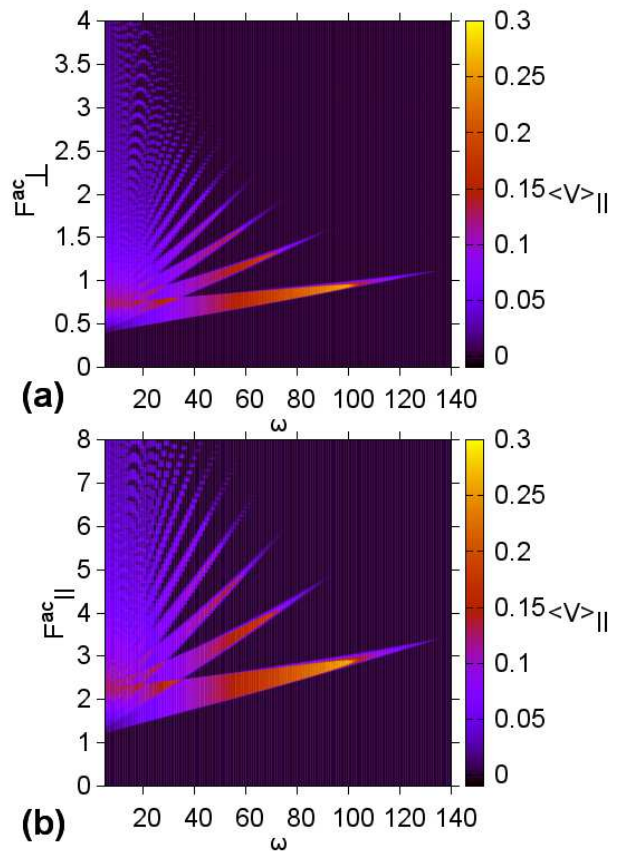}
\caption{
  {\bf Parallel ratchet velocity heat maps
    for parallel and perpendicular ac driving}:
  The magnitude of $\langle V\rangle_{\parallel}$
  in samples with $A_{p} = 1.5$ and $\alpha_m/\alpha_{d} = 3.042$.
  ({\bf a}) As a function of $F^{ac}_{\perp}$ and
  $\omega$ in units of $10^{-5}$ inverse simulation time steps,
  $\langle V\rangle_{\parallel}$ has clear
  tongue structures, and its magnitude is
  non-monotonic for varied frequency.
  ({\bf b}) $\langle V\rangle_{\parallel}$ as a function of
  $F^{ac}_{\parallel}$ and $\omega$ has similar features.  
}
\label{fig:7}
\end{figure}

\subsection{Frequency Dependence}
In Fig.~7(a) we plot a heat map of the
parallel ratchet velocity $\langle V\rangle_{\parallel}$
for perpendicular ac driving as a function of
$F^{ac}_{\perp}$ and $\omega$ in a sample with
$A_{p} = 1.5$ and $\alpha_{m}/\alpha_{d} = 3.042$.
The ratchet effect extends over a wider range
of $F^{ac}_{\perp}$ at low ac driving frequencies,
while for higher $\omega$  the
ratcheting regions form a series of tongues. The ratchet magnitude 
is non-monotonic as a function of $\omega$,
and in several regions it increases in magnitude with increasing
$\omega$, such as on
the first tongue where the maximum ratcheting effect
effect occurs near $\omega = 10^{-3}$ inverse simulation time steps
before decreasing again as $\omega$ increases further. We find
very similar ratchet behaviors for parallel ac driving,
as shown in Fig.~7(b) where we plot
$\langle V\rangle_{\parallel}$ as a function of
$F^{ac}_{\parallel}$ and $\omega$ and observe a series of tongues.

Our results show that skyrmions are an ideal system
in which to realize ratchet effects in the presence of asymmetric
substrates.
As expected, they undergo ratcheting when an ac drive is applied
parallel to the substrate asymmetry direction; however, the
skyrmions also exhibit a unique ratchet feature not found in overdamped
systems, which is the transverse ratchet effect.  Here, when
the ac drive is applied perpendicular to the substrate asymmetry direction,
the Magnus term curves the skyrmion orbits and drives them partially
parallel to the asymmetry, generating ratchet motion.
The asymmetric substrates we consider
could be created using methods similar to those employed to create
quasi-one-dimensional
asymmetric potentials in superconductors and related systems, such
as 
nanofabricated asymmetric thickness modulation,
asymmetric regions of radiation damage,
asymmetric doping,
or blind holes arranged in patterns containing density gradients.
Since the skyrmion ratchet effects
persist down to low frequencies,
it should be possible to directly image the ratcheting motion,
while for higher frequencies the existence of ratchet transport can be
deduced from transport studies.
Such ratchet effects offer a new method for controlling skyrmion
motion that could be harnessed in skyrmion applications.
We also expect that skyrmions
could exhibit a rich variety of other ratchet behaviors under different
conditions, such as more complicated substrate geometries,
use of asymmetric ac driving, or collective effects due to skyrmion-skyrmion
interactions.

\acknowledgments
This work was carried out under the auspices of the 
NNSA of the 
U.S. DoE
at 
LANL
under Contract No.
DE-AC52-06NA25396.
DR acknowledges support provided by the Center for Nonlinear Studies.

\end{document}